\documentclass[letter]{jpsj2}
\usepackage{graphicx}

\newcommand{\NFAO}{NdFeAsO$_{0.94}$F$_{0.06}$}
\newcommand{\Figs}[1]{Figs.~\ref{#1}}
\newcommand{\Fig}[1]{Fig.~\ref{#1}}

\title{Limits on the Superconducting Order Parameter in
  NdFeAsO$_{1-x}$F$_y$ from Scanning SQUID Microscopy }

\author{\textsc{Clifford W. Hicks},
 \textsc{Thomas M. Lippman},
 \textsc{Martin E. Huber}$^{1}$,
 \textsc{Zhi-An Ren}$^{2}$,
 \textsc{Zhong-Xian Zhao}$^{2}$,
 \textsc{Kathryn A. Moler}\thanks{E-mail: kmoler@stanford.edu}}

\inst{Geballe Laboratory for Advanced Materials, Stanford University, Stanford, California, 94305, USA
\\ $^{1}$Departments of Physics and Electrical Engineering,
     University of Colorado Denver, Denver, Colorado, 80217, USA \\
$^{2}$National Laboratory for Superconductivity, Institute of Physics and Beijing National Laboratory for
Condensed Matter Physics, Chinese Academy of Sciences, P.O. Box 603, Beijing 100190, P.R. China}

\abst{ Identifying the symmetry of the superconducting order parameter in the recently-discovered
ferro-oxypnictide family of superconductors, RFeAsO$_{1-x}$F$_{y}$, where $R$ is a rare earth, is a high
priority.  Many of the proposed order parameters have internal $\pi$ phase shifts, like the $d$-wave order
found in the cuprates, which would result in direction-dependent phase shifts in tunnelling. In dense
polycrystalline samples, these phase shifts in turn would result in spontaneous orbital currents and
magnetization in the superconducting state. We perform scanning SQUID microscopy on a dense polycrystalline
sample of \NFAO~ with $T_c=48$ K and find no such spontaneous currents, ruling out many of the proposed order
parameters.}

\kword{ferro-oxypnictides, pnictides, scanning SQUID microscopy, order parameter, NdFeAsO}

\begin{document}
\maketitle

The recently-discovered ferro-oxypnictide family of superconductors includes materials with transition
temperatures above 50 K~\cite{GdFeAs} and shows evidence that competing magnetism
plays a key role in the superconductivity~\cite{Chen, delaCruz, Liu}.
Determining the superconducting order parameter (OP) is key to understanding the interactions that induce
superconductivity, but the OP of the ferro-oxypnictide family of superconductors remains uncertain.
We report a phase-sensitive test of the symmetry of the OP using scanning magnetic microscopy of dense
polycrystalline samples. 

\begin{figure}[ptb]
\includegraphics[width=0.5\columnwidth]{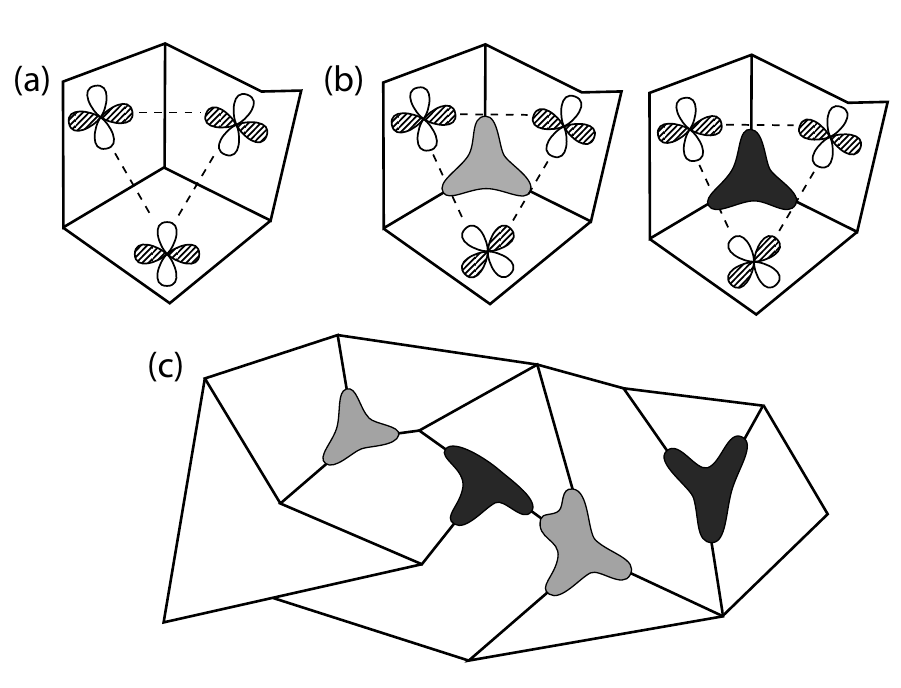}
\caption{\label{frustration} Spontaneous magnetic flux generated by various
configurations of grains. (a) Three coupled grains of a $d$-wave superconductor with no orbital frustration. The
orientation of the OP in each grain is indicated; shading indicates the sign of the OP and dashed lines
directions of strong intergrain tunnelling. (b) Coupled grains with orbital frustration. If $\lambda_J$ is
smaller than the grain size a half-flux-quantum ($\Phi_0/2$) vortex is generated, which can be positive (light)
or negative (dark). (c) A polycrystalline sample with several $\Phi_0/2$ vortices, which will tend
to couple antiferromagnetically. } 
\end{figure}

Grain boundaries form naturally occuring Josephson junctions that can carry supercurrents. It is now
well-known that the OP of the cuprate superconductors contains $\pi$ phase shifts associated with the $d$-wave
symmetry, and that a $\pi$ phase shift can result upon going around a closed path in a
polycrystalline sample; whether there is a $\pi$ shift depends on the relative lattice and interface
orientations of the grains along the loop~\cite{SigristRice}, for example as diagrammed in \Fig{frustration}.
$\pi$-loops result in orbital frustration and spontaneous currents, as demonstrated by observation of
half-integer flux quanta in tricrystal cuprate samples~\cite{Tsuei}. In polycrystalline cuprate samples there
is a finite density of $\pi$ loops, which, in well-connected samples ({\it i.e.} with the intergrain Josephson
penetration depth $\lambda_J$ comparable to or less than the grain size), results in complex patterns of
magnetization~\cite{KMSR}.

Proposals for the ferro-oxypnictide OP include extended-$s$ order, with a $\pi$ phase shift between the hole
and electron Fermi sheets~\cite{Mazin, Kuroki, WangLee, ChenZhang, Chubukov}, $d_{x^2-y^2}$\cite{Kuroki, Yao,
Weng, Qi}, $d_{xy}$~\cite{Si}, $p$~\cite{Lee, Qi}, $s+d$~\cite{Seo, LeeWu} and $s+id$~\cite{LeeWu}.
Phase-sensitive tests of the OP are
important: because of the multiple Fermi sheets, and because time-reversal symmetry-breaking (TRSB) OPs remain
a possibility, establishing the presence or absence of nodes does not definitively settle the OP symmetry.

Which OPs would result in orbital frustration in a polycrystalline sample? Any pure $d$ order will result in a
$\pi$ phase shift between $a$ and $b$ axis tunnelling, leading to frustration. In principle a 
TRSB component could reduce the degree of frustration: $d_{x^2-y^2}+id_{xy}$ order on a
radially symmetric band, and in a 2-D sample (the $c$-axes of the grains aligned), does not give
frustration. However, with the electron pockets, the ferro-oxypnictide Fermi surface is not radially
symmetric, and any such reduction would be minimal. $p$ order would result in frustration with or without a
TRSB component. $\pi$-shifted $s$ order in principle could result in a $\pi$ shift between $a$ and $c$ axis
tunnelling, if different, $\pi$-shifted, sections of the Fermi surface dominate $a$ and $c$ axis tunnelling.
Whether this is likely requires calculation. At present we must assume that an absence of spontaneous moments
does not rule out $\pi$-shifted $s$ order.

We have performed scanning SQUID imaging of a polycrystalline sample of nominal composition \NFAO~ grown by a
high-pressure synthesis method~\cite{renNd}. The superconducting transition onsets at 51 K, with a midpoint at
48 K and a 10\%-90\% width of 2.7 K. The grains are well-coupled: magneto-optical imaging and remnant
magnetization measurements on a sample from the same batch indicate a bulk critical current of $\approx 2000
$ A/cm$^2$ at 5 K~\cite{Yamamoto}.

\begin{figure*}[ht]
\includegraphics[width=0.85\textwidth]{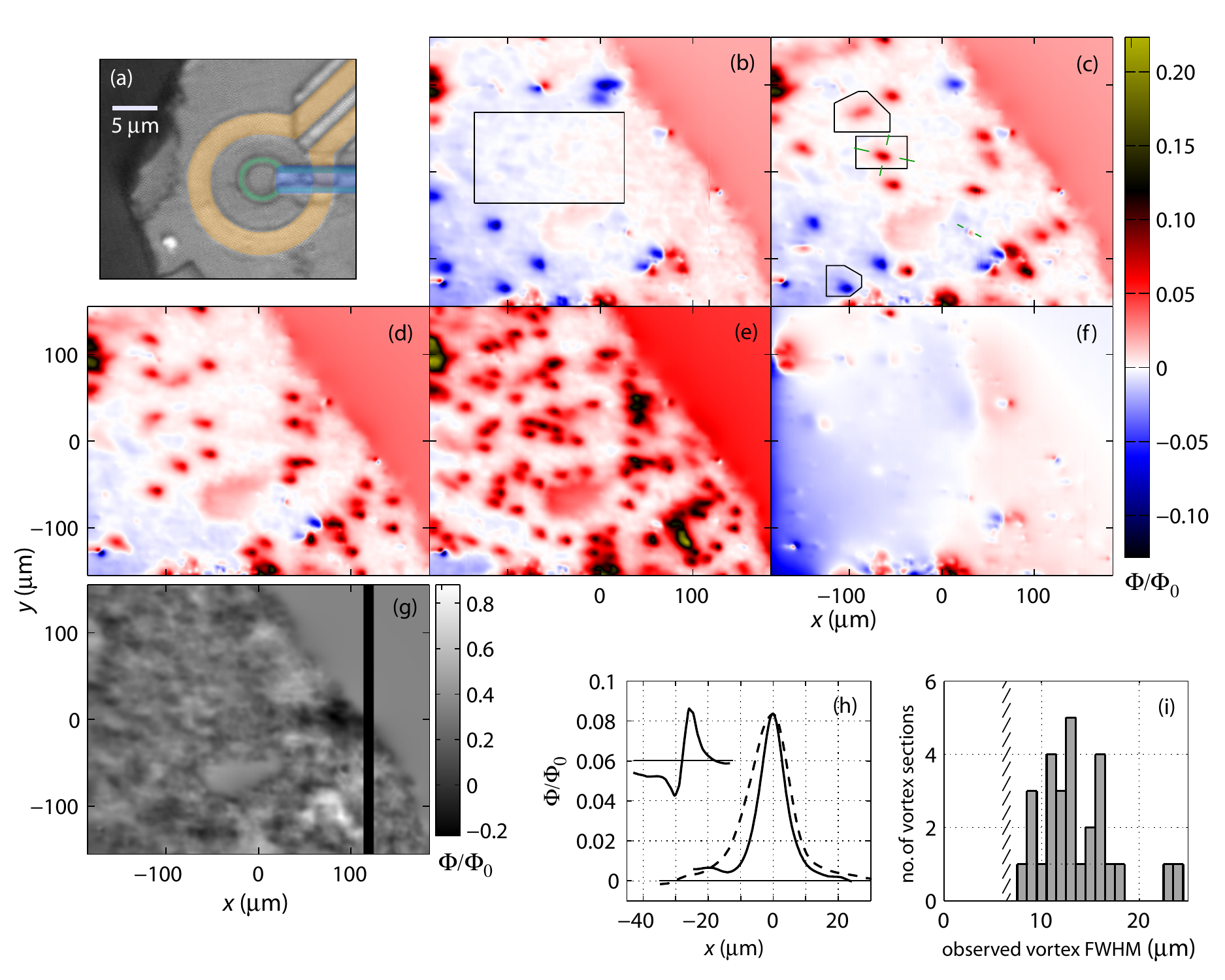}
\caption{\label{images} 
(a) Optical image of the front end of the SQUID; the field coil is highlighted in orange, the pick-up coil in
green, and the shield of the pick-up coil leads in blue.  (b-g) Images of the magnetic flux coupled into the
pick-up coil at sample temperature and applied field (b) 4.4 K, -3.3~$\mu$T; (c) 4.4, -2.7; (d) 4.4, -2.2; (e)
4.4, 0; (f) 55, 0; and (g) 4.4 K, +98~$\mu$T. An ambient $z$-axis field of $\approx3.3$~$\mu$T adds to these
fields.  The area indicated in (b) is used to analyze the background and the areas in (c) are integration
areas (see text).  (h) Cross sections along the green lines in (c): one section of a surface dipole and two of
a vortex.  (i) histogram of observed FWHMs of the isolated vortices in (c); two FWHMs per vortex, along the
narrow and long axes. Hatched area indicates the approximate resolution limit.}
\end{figure*}

Our SQUID is a niobium-based scanning susceptometer design~\cite{Huber}. \Fig{images}(a) contains an image of
the front end of the SQUID; magnetic flux is coupled into the 4.6 $\mu$m diameter pick-up coil (the inner
coil), the leads to which are shielded. In this SQUID a signal of 1 $\Phi_0=hc/2e=2.07\times10^{-15}$~Wb
corresponds to a mean $B_z$ in the pick-up coil of $\approx 0.125$~mT.  The larger loop around the pick-up
coil is a field coil; a measure of the local susceptibility can be obtained by applying a local field with
this coil and measuring the response in the pick-up coil.

We polished the sample to a shiny surface using Al$_2$O$_3$ polishing paper without any lubricant. In order to
allow comparison with the vacuum, we scanned an area going over the edge of the sample.
Our main results, scans of the sample cooled in different fields, are shown in \Fig{images}. At fields below
$\approx2$~$\mu$T  (panels (b)-(e)) individual vortices are clearly resolved. They appear
in different places on cooling in different fields, indicating that they are not frustration-induced
spontaneous moments. For confirmation that they are regular vortices, they can be integrated: after
subtracting a planar background, the three
vortices indicated in \Fig{images}(c) integrate to, from top to bottom, 1.05, 1.00 and -1.06 $\Phi_0$ (with a
5\% systematic uncertainty due to uncertainty in the effective pick-up coil area).


The 4.4 K scans 
reveal other features: several surface dipoles, clusters of vortices hinting at lumps
of a magnetic impurity phase, possibly beneath the surface, and a widespread mottled background. In the area
indicated in \Fig{images}(b), the root-mean-square amplitude of this background signal, after plane subtraction,
is 1.7 m$\Phi_0$. By lifting the SQUID slightly above the sample the sample can be heated while
maintaining the SQUID below its $T_c$; a $T=55$ K scan (\Fig{images}(f)) confirms the
presence of magnetic impurity phase. The surface dipoles also persist at 55 K while the mottled background
disappears at $T_c$.

\Fig{images}(g), a scan at $\approx100$~$\mu$T, makes clear the granular nature of the sample: vortices cluster
strongly in areas of weaker superconductivity.

\Fig{images}(h) shows sections of an isolated vortex and surface dipole from \Fig{images}(c). The surface
dipole provides a measure of the achieved imaging resolution: the peaks are separated by 5 $\mu$m. Assuming a
4.6 $\mu$m-diameter pick-up coil and a point-like dipole this indicates a scan height of the pick-up loop
above the sample surface of $\approx3$ $\mu$m. 2-D fits to the smallest dipoles in \Fig{images}(f) also
indicate a scan height of $\approx3$ $\mu$m.

\Fig{images}(i) is a histogram of the full-width half-maxima of the vortices in \Fig{images}(c). The narrowest
are $\approx8$ $\mu$m. With a 4.6 $\mu$m SQUID at a scan height of 3 $\mu$m a vortex in a sample with zero
penetration depth would appear with a FWHM of 6.2 $\mu$m; {\it i.e.} the 8 $\mu$m FWHMs are strongly
resolution-limited. However most of the vortices have observed FWHMs in the range 10--16 $\mu$m, and most have
visibly irregular shapes, suggesting that the actual vortices in the sample are spread out, with widths in the
range of microns.

\begin{figure}[htbp]
\includegraphics[width=0.5\columnwidth]{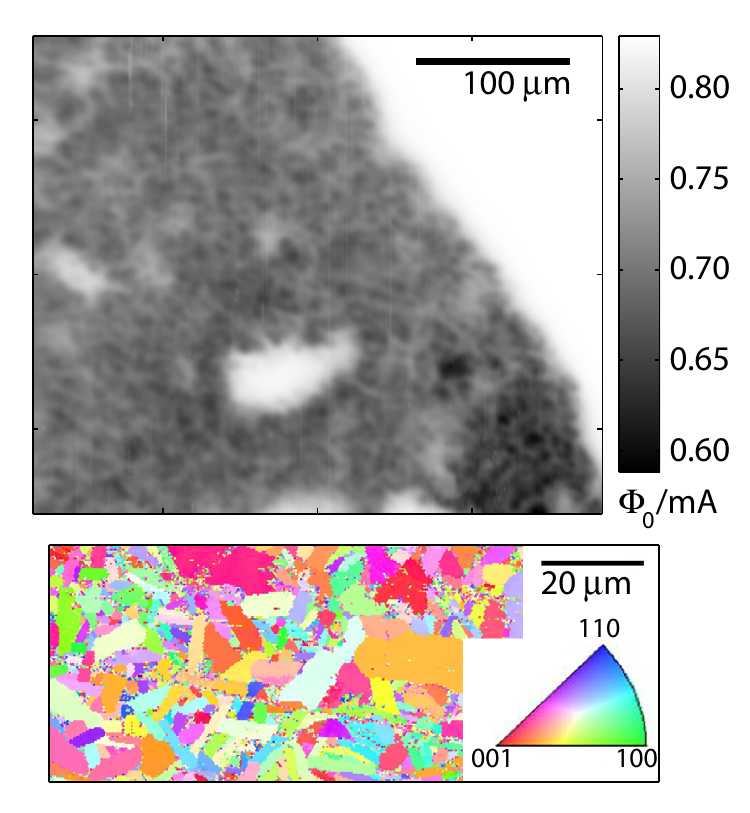}
\caption{\label{susc} 
Top: susceptibility scan at 4.4 K of the area imaged in \Fig{images}.
Bottom: electron backscatter diffraction scan of a polished surface of a piece of \NFAO from the same sample.}
\end{figure}

A susceptibility scan (\Fig{susc}) shows which areas are superconducting: over these areas the field coil is
partially shielded by the Meissner screening of the sample, reducing the field coil -- pick-up coil coupling
(measured in $\Phi_0$ of flux through the pick-up coil per mA of current in the field coil).
These areas appear dark in the figure.  The granular nature of the sample and areas of
non-superconducting phase are evident. (The lower right area appears most strongly superconducting,
however this is probably an artifact of topography allowing the SQUID closer to the sample.) \Fig{susc} also
shows an electron backscatter diffraction image, a technique which reveals crystal lattice orientation, of the
polished surface of a different piece of the same sample. The average grain diameter, in a circle
approximation, is 5.1 $\mu$m.

At first glance the mottled background observed in \Figs{images}(b)-(e) may resemble the complex magnetization
expected for orbital frustration in a polycrystalline sample. However, to the extent visible between vortices,
this background is identical in \Figs{images}(b)-(e), whereas frustration-related moments are polarizeable by
cooling in $\mu$T-scale applied fields~\cite{SigristRice}. Instead, the background is consistent with an
uncancelled in-plane field: as indicated in \Fig{images} we must apply $\approx-3.3$~$\mu$T to cancel the
$z$-axis component of the ambient field, and a comparable in-plane component can be expected. This would
result in in-plane vortices which would leak out near the surface of the inhomogeneous sample. Also, field
lines above the sample would be deflected upward and downward by the surface inhomogeneity. Both effects would
contribute to a mottled background signal.

\begin{figure}[htbp]
\includegraphics[width=0.5\columnwidth]{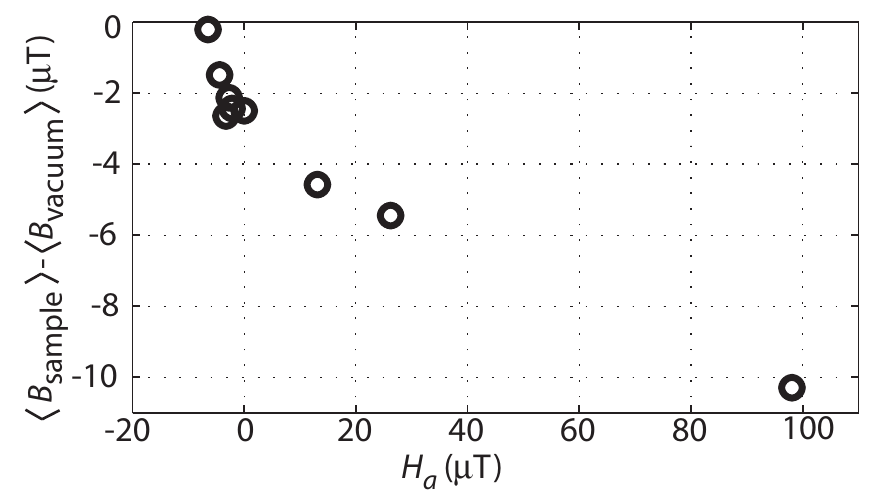}
\caption{ \label{diamag} From each scan, the difference between
the mean $B_z$ over the sample and in the corner farthest from the sample, against applied field (during
cooling and scanning) $H_a$. } 
\end{figure}

Polarizable moments would lead to the Wohlleben effect, a bulk paramagnetism against the field in which
the sample was cooled for fields $\lesssim100$~$\mu$T, and which has been observed for polycrystalline
cuprates~\cite{SigristRice}. To test for the Wohlleben effect in NdFeAsO$_{0.94}$F$_{0.06}$, we compare the
average signal over the sample with the signal beyond the sample edge (in all cases the sample was cooled and
scanned in the same field). The result, shown in \Fig{diamag}, indicates diamagnetism against sub-100~$\mu$T
cooling fields, consistent with an absence of polarizable moments. 

Qualitative examination of the scans and the absence of the Wohlleben effect indicate an
absence of orbital frustration in \NFAO. 

In the remainder of the paper, we use established modelling
techniques to show quantitatively that the signal that would emerge from orbital frustration would be larger
than the small mottled background field evident in \Fig{images}.
The observed vortex widths suggest a Josephson penetration depth $\lambda_J$ comparable to the grain
size, so spontaneous moments would not be well-isolated. Tightly-spaced moments would also tend to align
antiferromagnetically, further reducing the expected signal at the SQUID.
We estimate the expected signal from orbital frustration by modelling the grain interfaces as a long 1-D
Josephson junction, with a single $\lambda_J$, divided into $0$- and $\pi$-junction domains. A 1-D junction is
a reasonable approximation because the grain size is comparable to the system resolution. In the
narrow junction limit the phase change across the junction, $\phi(x)$, satisfies a sine-Gordon
equation,
\begin{equation} 
\frac{\partial^2 \phi}{\partial x^2} = \frac{1}{\lambda_J^2}\sin(\phi(x)+\theta(x)),
\end{equation}
where $\theta(x)$ is the position-dependent frustration phase (set here to 0 or $\pi$).

Two empirical estimates of the typical $\lambda_J$ for this sample 
are available.
\begin{equation}
\lambda_J = \left( \frac{\hbar c^2}{8 \pi e d j_c} \right)^2,
\end{equation}
where $d$ is the magnetic width of the junction and $j_c$ its critical current density. (The narrow junction
limit is $d\ll\lambda_J$.) $d=d_0+\lambda_1+\lambda_2$, where $d_0$ is the actual intergrain spacing and
$\lambda_1$ and $\lambda_2$ are the penetration depths of the two grains. The grain orientations being random
these will fall between $\lambda_{ab}$ and $\lambda_c$. $\lambda_{ab}$ has been measured at $\approx200$ nm in
$T_c\approx50$ K Sm- and Nd-based samples~\cite{Weyeneth, Khasanov, Drew}, and $\lambda_c/\lambda_{ab}\sim5$
has been measured in NdFeAsO$_{0.90}$F$_{0.10}$~\cite{Martin}. For $j_c \sim 2000$ A/cm$^2$ and $d \sim 2$
$\mu$m, $\lambda_J \sim 4$ $\mu$m is obtained.

The other estimate of $\lambda_J$ comes from the observed vortex widths. The soliton solution, for a vortex
within the junction, to the sine-Gordon equation is obtained by setting $\theta(x)$ to zero everywhere. To
extend this solution to above the sample we model it as a line of monopole sources an effective height $h$
beneath the pick-up coil, where $h$ is the actual scan height plus $\lambda$, and then integrate $B_z$ over
the pick-up coil area. $h$ can be estimated from the susceptibility scan shown in \Fig{susc}: over the sample
the field coil -- pick-up coil coupling is reduced by $\approx0.16$ $\Phi_0$/mA relative to in vacuum, which
would happen with the field and pick-up coils $\approx5$ $\mu$m above a hypothetical $\lambda=0$ plane.
Setting $h=5$ $\mu$m, observed vortex FWHMs of 10--16 $\mu$m indicate $\lambda_J$ in the range of 1--4 $\mu$m.

We simulate orbital frustration with a discretized junction 20,000 elements in length, divided into domains of
mean length $L=40$. In each domain $\theta(x)$ is set to $\pi$ with probability $P$, and zero otherwise.
$\phi(x)$ is obtained numerically as described in ref.~27. To simulate gradual cooling, the system was first
solved with $\lambda_J=200$, then $\lambda_J$ was reduced in steps to 10, taking the solution from the
previous step (with a small perturbation to disrupt unstable solutions) as the starting point for the next. 
$L$ and $\lambda_J$ are then scaled to lengths in microns, and the solutions are again extended to above the
sample by modelling as a line of monopole sources.

\begin{figure}[htbp]
\includegraphics[width=0.5\columnwidth]{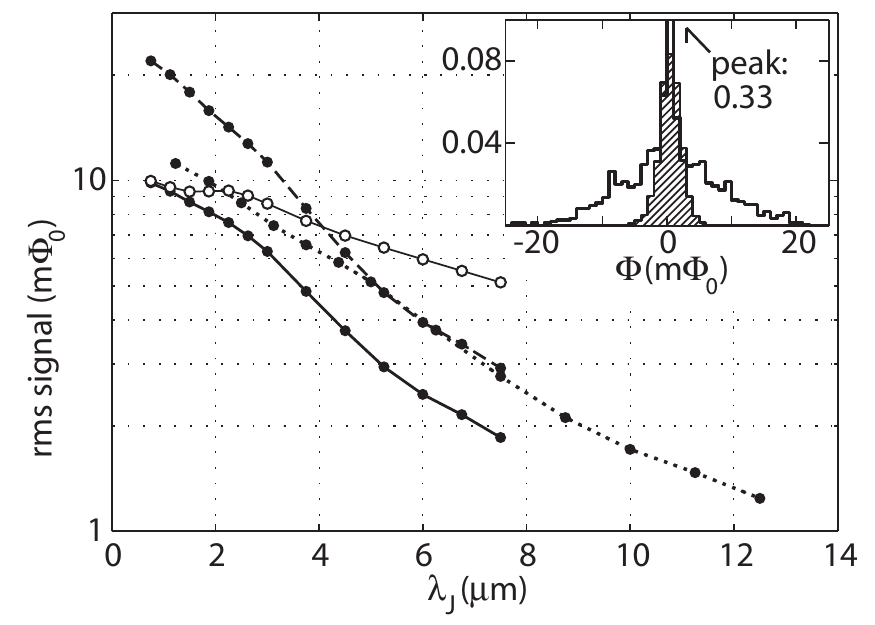}
\caption{\label{sim}
Root-mean-square expected SQUID signal from a frustrated system for various $\lambda_J$: results of the
simulation described in the text. Filled circles: $P=0.25$; 
solid line: $L, h = 3, 6$ $\mu$m, respectively; dashed: 3, 4 $\mu$m; dotted: 5, 6 $\mu$m.
Open circles: $P=0.50$, $L=3$ $\mu$m, $h=6$ $\mu$m. Inset: unshaded area:
histogram of expected signal (fraction per 1 m$\Phi_0$ window) for $L=3$ $\mu$m, $h=6$ $\mu$m, $\lambda_J=3$
$\mu$m, $P=0.25$; shaded area:
observed background in the area indicated in \Fig{images}(a) (fraction per window divided by
3).}
\end{figure}

The results of this simulation are shown in \Fig{sim}, and although it is an approximate model the simulation
shows that for a wide range of reasonable choices of $L$, $h$ and $\lambda_J$ a signal comparable to or larger
than the observed background would result: the fixed background is very unlikely to be obscuring an orbital
frustration signal. The inset compares the expected signal distribution for
the particular case $P=0.25$ and $L,h,\lambda_J=3$, 6, 3  $\mu$m, respectively, with the observed background
(after plane subtraction) in the area indicated in \Fig{images}(b).

We have demonstrated that there very likely are no $\pi$ phase shifts between tunnelling in different
directions in \NFAO, making $p$ and $d$ orders unlikely. $s$ order, $\pi$-shifted or not, and $s$+$d$ order
where the $d$ component is small are not ruled out by our result.

This project was supported by the U.S. Department of Energy (DE-AC02-76SF00515).
We thank David Larbalestier, Alex Gurevich, and Doug Scalapino for
useful discussion. We also thank Fumitake Kametani and David Larbalestier for providing the
EBSD image.


\begin{thebibliography}{99}

\bibitem{GdFeAs}Several $T_c>50$ K materials are now known, for example in Z.A. Ren, Wei Lu, Jie Yang, Wei Yi,
X.L. Shen, Cai Zheng, G.C. Che, X.L. Dong, L.L. Sun, Fang Zhou and Z.X. Zhao: Chin. Phys. Lett. {\bf 25}
(2008) 2215.

\bibitem{Chen}G.F. Chen, Z. Li, D. Wu, G. Li, W.Z. Hu, J. Dong, P. Zheng, J.L. Luo and N.L. Wang: Phys. Rev.
Lett. {\bf 100} (2008) 247002.

\bibitem{delaCruz}C. de la Cruz, G. Huang, J. W. Lynn, Jiying Li, W. Ratcliff II, J.L. Zaretsky, H.A. Mook,
G.F. Chen, J.L. Luo, N.L. Wang and Pengcheng Dai: Nature {\bf 453} (2008) 899.

\bibitem{Liu}R.H. Liu, G. Wu, T. Wu, D.F. Fang, H. Chen, S.Y. Li, K. Liu, Y.L. Xie, X.F. Wang, R.L. Yang, L.
Ding, C. He, D.L. Feng, X.H. Chen: Phys. Rev. Lett. {\bf 101} (2008) 087001.

\bibitem{SigristRice}M. Sigrist and T.M. Rice: Rev. Mod. Phys. {\bf 67} (1995) 503.

\bibitem{Tsuei}C.C. Tsuei, J.R. Kirtley, C.C. Chi, Lock See Yu-Jahnes, A. Gupta, T. Shaw, J.Z. Sun and M.B.
Ketchen: Phys. Rev. Lett. {\bf 73} (1994) 593.

\bibitem{KMSR}J.R. Kirtley, A.C. Mota, M. Sigrist and T.M. Rice: J. Phys.: Condens. Matter {\bf 10} (1998)
L97.

\bibitem{Mazin}I.I. Mazin, D.J. Singh, M.D. Johannes and M.H. Du: Phys. Rev. Lett. {\bf 101} (2008) 057003.

\bibitem{Kuroki}K. Kuroki, S Onari, R. Arita, H. Usui, Y. Tanaka, H. Kontani and H. Aoki: Phys. Rev. Lett.
{\bf 101} (2008) 087004.

\bibitem{WangLee}Fa Wang, Hui Zhai, Ying Ran, A. Vishwanath and D.H. Lee: cond-mat/08070498.

\bibitem{ChenZhang}W.Q. Chen, K.Y. Yang, Yi Zhou and F.C. Zhang: cond-mat/08083234.

\bibitem{Chubukov}A.V. Chubukov, D.V. Efremov and I. Eremin: Phys. Rev. B {\bf 78} (2008) 134512.

\bibitem{Yao}Z.J. Yao, J.X. Li and Z.D. Wang: cond-mat/08044166.

\bibitem{Weng}Z.Y. Weng: cond-mat/08043228.

\bibitem{Qi}X.L. Qi, S. Raghu, C.X. Liu, D.J. Scalapino, S.C. Zhang: cond-mat/08044332.

\bibitem{Si}Q.M. Si and E. Abrahams: Phys. Rev. Lett. {\bf 101} (2008) 076401.

\bibitem{Lee}P.A. Lee and X.G. Wen: cond-mat/08041739.

\bibitem{Seo}K. Seo, B.A. Bernevig and Jiangping Hu: Phys. Rev. Lett. {\bf 101} (2008) 206404.

\bibitem{LeeWu}W.C. Lee, S.C. Zhang and C.J. Wu: cond-mat/08100887.

\bibitem{renNd}Z.A. Ren, Jie Yang, Wei Lu, X.L. Shen, Z.C. Li, G.C. Che, X.L. Dong, L.L. Sun, Fang Zhou and
Z.X. Zhao: Europhys. Lett. {\bf 82} (2008) 57002.

\bibitem{Yamamoto}A. Yamamoto, A.A. Polyanskii, J. Jiang, F. Kametani, C. Tarantini, F. Hunte, J. Jaroszynski,
E.E. Hellstrom, P.J. Lee, A. Gurevich, D.C. Larbalestier, Z.A. Ren, J. Yang, X.L. Dong, W. Lu and Z.X. Zhao:
Supercond. Sci. Technol. {\bf 21} (2008) 095008.

\bibitem{Huber}M.E. Huber, N.C. Koshnick, H. Bluhm, L.J. Archuleta, T. Azua, P.G. Bj\"{o}rnsson, B.W. Gardner,
S.T. Halloran, E.A. Lucero and K.A. Moler: Rev. Sci. Instrum. {\bf 79} (2008) 053704.

\bibitem{Weyeneth}S. Weyeneth, U. Mosele, N.D. Zhigadlo, S. Katrych, Z. Bukowski, J. Karpinski, S. Kohout, J.
Roos and H. Keller: cond-mat/08061024.

\bibitem{Khasanov} R. Khasanov, H. Luetkens, A. Amato, H.H. Klauss, Z.A. Ren, Jie Yang, Wei Lu and Z.X. Zhao:
Phys. Rev. B {\bf 78} (2008) 092506.

\bibitem{Drew}A.J. Drew, F.L. Pratt, T. Lancaster, S.J. Blundell, P.J. Baker, R.H. Liu, G. Wu, X.H.
Chen, I. Watanabe, V.K. Malik, A. Dubroka, K.W. Kim, M. Roessle and C. Bernhard: Phys. Rev. Lett. {\bf 101}
(2008) 097010.

\bibitem{Martin}C. Martin, R.T. Gordon, M.A. Tanatar, M.D. Vannette, M.E. Tillman, E.D. Mun, P.C. Canfield,
 V.G. Kogan, G.D. Samolyuk, J. Schmalian and R. Prozorov: cond-mat/08070876.

\bibitem{KMS}J.R. Kirtley, K.A. Moler and D.J. Scalapino: Phys. Rev. B {\bf 56} (1997) 886.



\end{thebibliography}
\end{document}